\documentclass{article}
\usepackage{spconf,amsmath,graphicx}
\usepackage[english]{babel}

\usepackage[utf8]{inputenc} 
\usepackage{url}
\usepackage{multirow}
\usepackage{multicol}
\usepackage{enumitem}
\usepackage{subcaption}
\usepackage{dsfont}
\usepackage[font=small]{caption}
\usepackage{tabularx}
\usepackage{color}  
\usepackage{hyperref}  

\usepackage[nolist]{acronym}
\begin{acronym}
    \acro{MIR}{Music Information Retrieval}
    \acro{STFT}{Short Time Fourier Transform}
    \acro{CNN}{Convolutional Neural Network}
    \acrodefplural{CNN}{Convolutional Neural Networks}
    \acro{DNN}{Deep Neural Network}
    \acro{CQT}{Constant-Q transform}
    \acro{MAP}{Mean Average Precision}
    \acro{nDCG}{Normalized Discounted Cumulative Gain}
\end{acronym}

\title{Learning to rank music tracks using triplet loss}

\name{Laure Prétet$^{\star \dagger}$ \qquad Gaël Richard$^{\star}$ \qquad Geoffroy Peeters$^{\star}$ }
\address{$^{\dagger}$ Creaminal, Paris, France \\
    $^{\star}$ LTCI, Télécom Paris, Institut Polytechnique de Paris, France}
    
\begin{document}
\ninept

\maketitle

\begin{abstract}
Most music streaming services rely on automatic recommendation algorithms to exploit their large music catalogs.
These algorithms aim at retrieving a ranked list of music tracks based on their similarity with a target music track. 
In this work, we propose a method for direct recommendation based on the audio content without explicitly tagging the music tracks. To that aim, we propose several strategies to perform triplet mining from ranked lists.
We train a Convolutional Neural Network to learn the similarity via triplet loss.
These different strategies are compared and validated on a large-scale experiment against an auto-tagging based approach. 
The results obtained highlight the efficacy of our system, especially when associated with an Auto-pooling layer. 
\end{abstract}







\section{Introduction}
\label{sec:intro}

Many domains, such as music streaming services, make use of large music catalogs.
To organize these tracks, it is necessary to provide efficient retrieval mechanisms. 
While browsing by tags (genre, mood, instrumentation) can be efficient for small-scale catalogs, it does not provide an efficient retrieval mechanism at scale.
This is why most music streaming services rely on music recommendation.
For this purpose, an algorithm is used to retrieve a ranked list of music tracks based on their similarity with a target music track. 
Provided a similarity metric is defined, the music ranking problem reduces to a \textit{music similarity problem}.
This task has been the subject of many publications (see \cite{Urbano2013EvaluationSimilarity} for an overview) and can be tackled in different ways.

In \textit{Collaborative filtering recommendation}, two music tracks can be considered similar if they are listened to by the same audience \cite{Oord2013DeepRecommendation}~\cite{Chen2001AInterests}.
Obviously, if no one has ever listen to a music track (for example a new track in the catalog), it can not be recommended.
This is known as the cold start problem \cite{Schein2002MethodsRecommendations}.
However, when applicable, collaborative filtering has proved to provide very good results for recommendation.

In \textit{Tag-based recommendation}, a tag-based similarity measure can be designed to compare music tracks based on their respective tags.
Tags can be manually annotated (such as in Pandora \cite{Clifford2007PandorasTrip}), crowd-sourced (such as in Last.fm), or automatically inferred from the audio content (auto-tagging case \cite{Choi2016AutomaticNetworks}).

In \textit{Direct recommendation}, it is possible to compute directly a distance between two music tracks based on their audio similarity.
For example, in one of the pioneer works \cite{Aucouturier2002FindingSame}, track MFCCs were represented by a Gaussian Mixture Model and a Kullback-Leibler divergence was used to compare two tracks. 
Since these methods use a costly pairwise comparison, they cannot scale to large catalogs.

In this work, we propose a method for \textit{direct recommendation} based on the audio content.
To this aim, we suppose we have access to a large music catalog professionally annotated with tags.
We also assume we are given a function $\mathcal{S}$ which allows to compute a similarity score between two sets of tags. 
For a given target track, $\mathcal{S}$ allows us to retrieve similar tracks based on their tags.
Our goal is to reproduce the similarity ranking given by $\mathcal{S}$ \textit{without explicitly tagging the tracks}. 
We denote our approximation by $\mathcal{\hat{S}}$.
For a given target track, $\mathcal{\hat{S}}$ allows us to retrieve similar tracks based on audio directly.
This allows to skip the long and expensive manual tagging step.
In this work, we tackle this task using deep learning.
We train a \ac{CNN} with triplet loss to learn a projection of the audio signal such that the proximity between two projected music tracks accounts for their similarity $\mathcal{S}$.
We compare this similarity to the one obtained by a \ac{CNN} trained to estimate automatically the related tags which are then used in $\mathcal{S}$.

\textbf{Paper contributions.}
The three main contributions of our work are the following.
First, we propose several strategies to perform triplet mining from ranked lists. 
This allows to apply a triplet loss to the relative music similarity problem. 
Second, we compare and validate these different strategies on a large-scale experiment against the auto-tagging based approach. 
Third, we demonstrate the efficacy of the recently proposed Auto-pooling layer \cite{McFee2018AdaptiveDetection} for a music task.

\textbf{Paper organization.}
In section \ref{sec:related}, we review works related to ours.
In section \ref{sec:method}, we describe our proposed method to mine triplets from ranked lists. 
In section \ref{sec:eval}, we evaluate the proposed approach, along with an auto-tagger baseline, on a music retrieval task. 
In section \ref{sec:conclusion}, we draw the conclusions of our results.


\vspace{-3mm}

\section{Related work} 
\label{sec:related}


\subsection{Music auto-tagging}

In \ac{MIR}, we call \textit{music auto-tagging} the task of predicting tags directly from audio signals. 
Tags are keywords that describe a music track in terms of genre, mood, instrumentation or any other high-level attributes. 

Traditional approaches for music auto-tagging rely on the extraction of handcrafted features to feed a classifier (linear or non-linear) \cite{Eck2008AutomaticRecommendation} \cite{Costa2017AnSpectrograms}. 
Recently, deep learning approaches have allowed to learn the features directly from the data (waveforms or spectrograms), leading to improved performances \cite{Dieleman2014End-to-endAudio} \cite{Pons2018End-to-endScale} \cite{Pons2017TimbreNetworks} \cite{Choi2016AutomaticNetworks}.
Some of these systems have proven their capacity to learn useful information from audio \cite{Dieleman2014End-to-endAudio} \cite{Choi2016ExplainingClassification}. 
Therefore, we take inspiration from those for our \ac{CNN} architecture, but we use this \ac{CNN} to learn music similarity instead of tags.

\subsection{Learning to rank}

The task of retrieving items in a collection and to sort them by relevance arises in a variety of domains. 
In early works, Weston \cite{WestonJasonYeeHectorWeiss2013LearningLoss.} and Usunier \cite{Usunier2009RankingClassification.} proposed loss functions capable of optimizing the top of ranked lists in matrix factorization and information retrieval contexts.
Another common approach is to learn a similarity notion from classes. 
In this case, it is assumed that the similarity of items within a same class should be higher than the similarity of items from different classes. 
As a result, the model's recommendations are associated to a binary relevance score: the recommended item does or does not belong to the expected class. 
Such problems have been studied for example by Weinberger and Saul~\cite{Weinberger2009DistanceClassification} and Hoffer and Ailon \cite{Hoffer2015DeepNetwork}, who employed \acp{CNN} with innovative loss functions to perform higher-accuracy image classification. 
Today, a widely used loss for similarity learning is the \textit{triplet loss}, as introduced by \cite{Weinberger2009DistanceClassification} and used by Schroff and Philbin \cite{Schroff2015FaceNet:Clustering} for face recognition. 
In their work, they use the triplet loss to force the \ac{CNN} to learn a projection of the image data such that the projections of images of the same person will be pulled together, and the ones from different persons will be pushed apart.
The distance employed to compare the projections of the data is usually the Euclidean distance, squared Euclidean distance, or the cosine distance \cite{Hermans2017InRe-Identification}.

In our problem (learning a similarity from ranked lists), there are however no classes, nor binary relevance labels for each query-document pair. 
Such a problem has already been addressed outside the music case.
For example, Mcfee and Lanckriet \cite{Mcfee2010MetricRank} propose to use a listwise loss function to learn text recommendation from ranked lists. 
Wang et al. \cite{Wang2014LearningRanking} propose to use the triplet loss to learn a ranking of similar images. 
Our proposal takes inspiration from the work of \cite{Wang2014LearningRanking} but for the music case.

\vspace{-0.3cm}
\subsection{Music similarity}

The literature on music similarity is vast (see Wolff and Weyde~\cite{Wolff2014LearningRatings} for an overview).
For example, Slaney et al. \cite{Slaney2008LearningSimilarity} propose a set of linear transforms to embed tracks into an Euclidean metric space and evaluate them on a nearest neighbor task for album, artist and blog recognition. 
Tag-based approaches to metric learning include a method by Weston et. al. \cite{Weston2011Large-ScaleSpaces} to project both audio features and music tags into a shared embedding space. 
Wolf and Weyde~\cite{Wolff2014LearningRatings} insist on modeling \textit{relative} music similarity (rankings) rather than absolute similarity ratings, in order to avoid consistency issues due to subjective user ratings.
Following this idea, Lu et al. \cite{Lu2017DeepLearning} employ a \ac{CNN} with an improved triplet loss to predict relative music similarity.
However \cite{Lu2017DeepLearning} do not propose any mining strategies from ranked lists\footnote{This is because these two last studies train their model using the similarity triplet annotations provided in the MagnaTagATune dataset \cite{Law2009EvaluationTagging}, which has only partial similarity information, annotated by non-expert users.}. 
Our work takes inspiration from \cite{Lu2017DeepLearning} but proposes a mining strategy from ranked lists. 



\vspace{-0.3cm}
\section{Proposed method}
\label{sec:method}

\subsection{Problem definition}
\label{ssec:defs}

Let $D = \{t_1, \ldots, t_N\}$ be a set of $N$ tracks annotated with a taxonomy of $m$ tags. 
The problem addressed in this study is the following: given a query track $t$, compute a ranked list of tracks from the dataset $D$ ordered by descending similarity to $t$. 
Let $\mathcal{S}$ be an oracle similarity function that, given two sets of tag likelihoods $t_1 \in [0,1]^m$ and $t_2 \in [0,1]^m$, returns a similarity score $\mathcal{S}(t_1, t_2) \in \mathds{R}_{+}$.

For any $t \in D$, let $R^{\mathcal{S}}(t) = [r_1(t), \ldots, r_{N-1}(t)]$ be the ordered list of the other tracks of $D$ ranked by decreasing similarity according to the function $\mathcal{S}$. 
This is the \textbf{ground truth}, against which our system will be evaluated.
For any $t \in D$, let $R^{\mathcal{\hat{S}}}(t)~=~[\hat{r}_1(t), \ldots, \hat{r}_{N-1}(t)]$ be the estimated list of recommended tracks made by the system for the target track $t$. 

In this paper, we formulate the music ranking problem as a nearest neighbor search problem in a $d$-dimensional Euclidean space. 
A model is trained to define a specific \textit{embedding space} (a projection of the data) in which the Euclidean distance allows to retrieve tracks with the same ranking as with the oracle similarity function $\mathcal{S}$.

\vspace{-0.3cm}    
\subsection{Mining triplets from ranked lists}
\label{ssec:mining}

We denote by $f(t) \in \mathds{R}^d$ the embedding of the track $t$.
$f$ is obtained by training a \ac{CNN} using a triplet loss. 

This loss takes as input a triplet of tracks that consists of an anchor \textit{a}, a positive example \textit{p} and a negative example \textit{n}. 
The \ac{CNN} outputs the embedding vectors of those, \textit{f(a), f(p)} and \textit{f(n)} respectively.
The triplet is created such that the positive is more similar to the anchor than the negative according to the ground truth $\mathcal{S}$.
The triplet loss then compares the squared Euclidean distances between the three embedding vectors and ensures that the same condition is respected in the embedding space:
\begin{equation*}
\small
    \mathcal{L}(a,p,n)= \max(||f(a)-f(p)||^2_2 - ||f(a)-f(n)||^2_2 + \alpha, 0).
\end{equation*}
In this expression, $\alpha \in \mathds{R}_+^*$ is a margin parameter that enforces a minimal distance between the positive and negative pairs.

To train such a system, it is necessary to prepare the data in the form of triplets $(a,p,n)$.
Let $t$ be a target track and $R^{\mathcal{S}}(t) = [r_1(t), \ldots, r_{N-1}(t)]$ the associated reference ranking.
We define training triplets by using the track $t$ as the anchor and by mining a positive and a negative element from $R^{\mathcal{S}}(t)$. 
A triplet is considered \textit{valid} if the index of the positive element is lower than the one of the negative element: $(a,p,n) = (t, r_i(t), r_j(t))$ \hspace{0.2 cm} $\forall i < j$. 

\begin{figure}[thpb]
   \centering
    \includegraphics[scale=0.4]{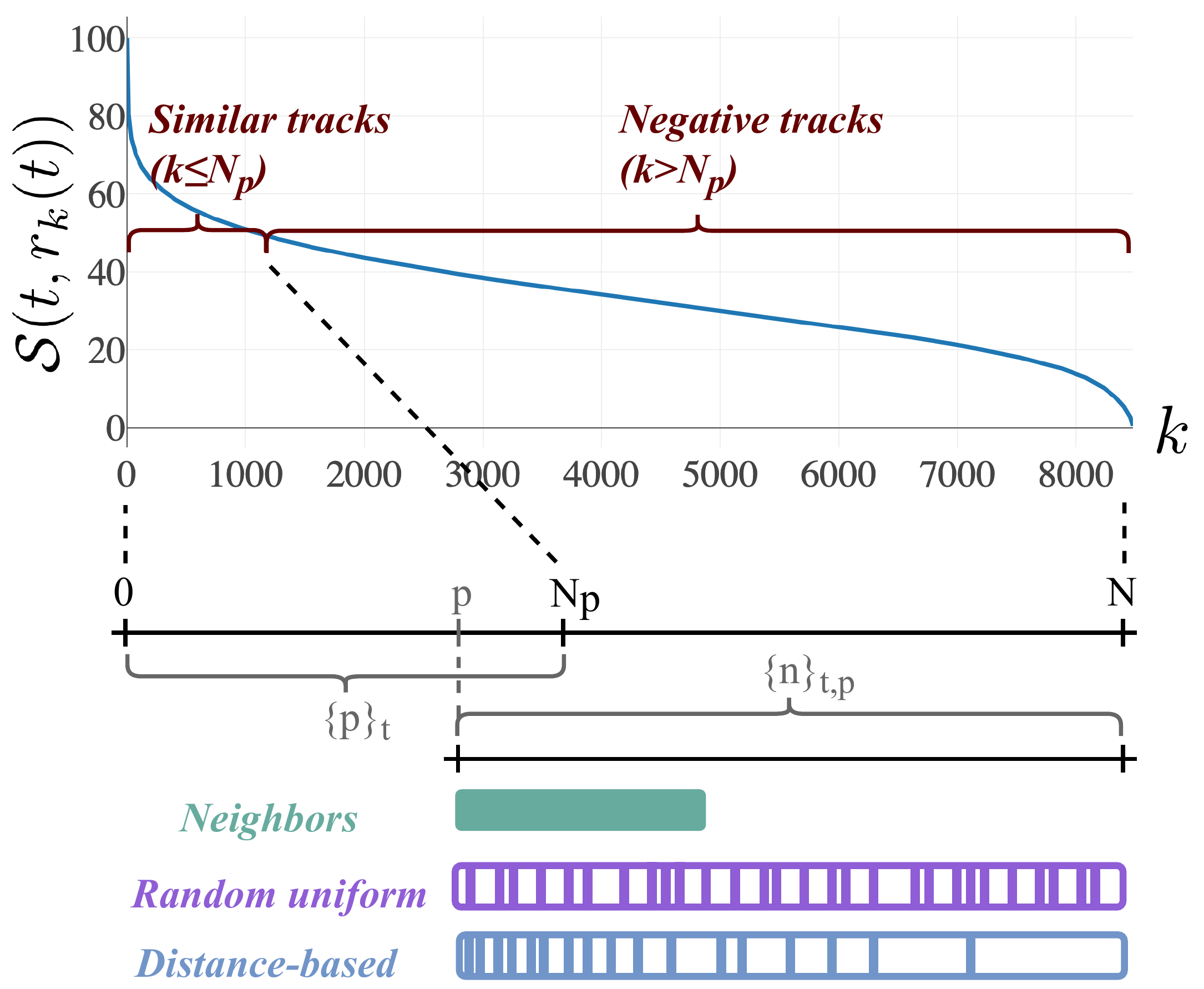}
    \caption{Illustration of the three proposed triplet mining strategies. Top: Similarity scores in a typical similarity ranking (in percentage), as a function of the rank in the list $R^{\mathcal{S}}(t)$. Bottom: Three strategies to mine negative samples for a given anchor-positive pair $(t,p)$.}
   \label{fig:sim}
 \end{figure}

In practice, for large datasets, it is infeasible to use all valid triplets for training, because their number grows cubically with $N$.
Additionally, all triplets may not be useful for training. 
Figure \ref{fig:sim} (top) shows an illustration of the average similarity scores $[\mathcal{S}(t,r_1(t)), \ldots, \mathcal{S}(t,r_{N-1}(t))]$ for each track $t \in D$.
The curve shows that a few first tracks in the ranking are very similar to their target, while most tracks in the dataset are actually irrelevant: their similarity score is lower than 50\%.
Therefore, after a certain rank in $R^{\mathcal{S}}(t)$, mining positive samples does not make sense.

Given these observations, we limit the set of possible positives per anchor to the $N_p$ first elements of the reference ranking: $p \in [r_1(t), \ldots, r_{N_p}(t)]$.
We also limit the overall number of negatives per anchor-positive pair to $N_n$. 
Then, to select the $N_n$ negative examples for a given anchor-positive pair \textit{$(t, r_i(t))$}, three different strategies are tested (see Figure \ref{fig:sim}, bottom):

\begin{itemize}[leftmargin=4mm, parsep=0cm, itemsep=0cm, topsep=0cm]
    \item \textbf{Neighbors:} The negatives are the $N_n$ elements in $R^{\mathcal{S}}(t)$ that come directly after the positive:
    $n \in [r_{i+1}(t), \ldots,r_{i+1+N_n}(t)]$.
    \item \textbf{Random uniform:} The negatives are sampled uniformly among the full $R^{\mathcal{S}}(t)$ list after the positive: $n \in \{r_j(t)\}_{j > i}$ s.t. $P(n) \sim \mathcal{U}(1/(N-i+1))$.
    \item \textbf{Distance-based} (inspired by \cite{Manmatha2017SamplingLearning} and \cite{Wang2014LearningRanking}): The negatives are sampled among the full $R^{\mathcal{S}}(t)$ list after the positive with a probability that is proportional to its similarity with the anchor: $n \in \{r_j(t)\}_{j > i}$ s.t. $P(n) \propto \mathcal{S}(t,n)$.
\end{itemize}
 
This way, in all three variants, the set of triplets can be pre-selected offline and we do not mine triplets during training.
The resulting total number of triplets is $N\times N_p\times N_n$.

\subsection{Auto-pooling}

In usual \acp{CNN} applied to audio signals, the time dimension of the audio signal is progressively discarded by a succession of max-pooling layers.
This implies a strong assumption related to how information over time is processed: we only keep the maximum activation over successive time frames.
Other choices have been made in the past such as the combination of Mean, Max and L2 Pooling~\cite{Dieleman2014RecommendingLearning}.
A very elegant formulation has been proposed by McFee et al. \cite{McFee2018AdaptiveDetection} with the Auto-Pooling layer, which allows to interpolate between several pooling operators (such as min-, max-, and average-pooling) via a learned parameter.
Auto-pooling has provided very good results for an audio event detection task. 
To our knowledge, it has not been used for music-related tasks.
We do this here for the task of music similarity.


\subsection{System architectures}

In the following, the input representation used for all our architectures is the \ac{CQT}.
For each track, we compute a \ac{CQT} of 96 bins (12 bins/octave) with $f_{min}$=32.70~ Hz, and a hop size of 1024 at 44.1~kHz.
We then convert it to power amplitude and log-scale the magnitudes.
The input of our \ac{CNN} is a patch of 512 CQT frames (11.88s). 
This duration was chosen as a good compromise between memory efficiency and sufficient musical context.
Since the annotations of our dataset are at the track level (see Section~\ref{sec:dataset}), 
we randomly select several of these (96$\times$512) patches to represent a given track and assume that the annotations apply to each patch. 

At test time, each track is represented by 8 randomly selected patches.
When the network is used for auto-tagging, we pass each of them through our network to obtain the estimated likelihood vector.
We then simply use the average vector over the 8 tag estimated likelihood vectors.
When the network is used for embedding, we compute the average embedding vector over the 8 embedding vectors.

All models have been implemented using Keras with an Adam optimizer, a batch size of 42 patches and early stopping. 

\begin{table}[t!] 
\small
    \centering
    \begin{tabular}{|c|c|c|}
        \hline
        \textbf{AT Baseline} & \textbf{TL} & \textbf{TL Autopool} \\
        \hline
         \multicolumn{3}{|c|}{CQT \textit{(input: F=96,T=512,C=1)}} \\ \hline
         \multicolumn{2}{|c|}{Conv2D (3,3)$\times$128} & Conv2D (3,3)$\times$64 \\ \hline 
         \multicolumn{2}{|c|}{MP (2,4) \textit{(F=48,T=128,C=128)}} & MP (2,2) \textit{(F=48,T=256,C=64)}\\ \hline
         \multicolumn{2}{|c|}{Conv2D (3,3)$\times$256} & Conv2D (3,3)$\times$128 \\ \hline 
         \multicolumn{2}{|c|}{MP (2,4) \textit{(F=24,T=32,C=256)}} & MP (2,2) \textit{(F=24,T=128,C=128)} \\ \hline
         \multicolumn{2}{|c|}{Conv2D (3,3)$\times$512} & Conv2D (3,3)$\times$256 \\ \hline 
         \multicolumn{2}{|c|}{MP (2,4) \textit{(F=12,T=8,C=512)}} & MP (2,2) \textit{(F=12,T=64,C=256)} \\ \hline
         \multicolumn{2}{|c|}{Conv2D (3,3)$\times$1024} & Conv2D (3,3)$\times$512 \\ \hline 
         \multicolumn{2}{|c|}{MP (3,3) \textit{(F=4,T=2,C=1024)}} & MP (2,2) \textit{(F=6,T=32,C=512)} \\ \hline
         \multicolumn{2}{|c|}{Conv2D (3,3)$\times$2048} & Conv2D (3,3)$\times$1024 \\ \hline 
         \multicolumn{2}{|c|}{MP (4,2) \textit{(F=1,T=1,C=2048)}} & MP (6,1) \textit{(F=1,T=32,C=1024)} \\ \hline
          &  & FC ($d$) \textit{(F=1,T=32,C=$d$)} \\ \hline 
          FC \textit{($m$)} & FC \textit{($d$)} & Autopool (1,32) \textit{($d$)} \\ \hline
    \end{tabular}
    \caption[]{Details of the three architectures used. In italic are the output sizes of the Max-Pooling (MP) or Fully-Connected (FC) layers.}
    \label{tab:archi}
\end{table}

\vspace{0.3cm}


{\bf Baseline system: Auto-tagger (AT)}: The similarity function $\mathcal{S}$, used for ranking, relies on tag likelihood annotations.
A naive approach is therefore to automatically estimate these tag likelihoods from the audio and then apply $\mathcal{S}$ directly to the estimated likelihood vectors.
Auto-tagging is a multi-label classification problem (output activations are sigmoids, loss defined as the sum of binary cross-entropies).
Preliminary experiments have shown that VGG-like architectures \cite{Choi2016AutomaticNetworks} were more suited to our dataset than musically-motivated architectures \cite{Pons2017TimbreNetworks}. 
Thus, we reproduce the \textit{FCN-5} architecture proposed by Choi et al. ~\cite{Choi2016AutomaticNetworks}. 
We adapt it to the shape of our inputs (96$\times$512) and outputs ($m$ tags).
While the ground-truth annotations are likelihoods in $[0,1]^m$, we train the system with binarized outputs $\{0,1\}^m$.
This system is referred to as \textit{AT Baseline} in the rest of the paper.
We give its architecture in Table~\ref{tab:archi}, column 1 and provide the details (dropout, activations) in our code.
We train it with a learning rate of $10^{-4}$. 

\vspace{0.3cm}


{\bf Triplet loss system (TL)}: Our objective here is to estimate directly an embedding such that applying the Euclidean distance between the embeddings of two tracks $t_1, t_2$ mimics $\mathcal{S}(t_1,t_2)$.
The network we use to compute the embedding is similar to the \textit{AT Baseline} one, but it differs in the output layer.
The last fully-connected layer has now $d$ units (the dimension of the embedding space) instead of the $m$ units, and has linear activations instead of $m$ sigmoid activations (see Table \ref{tab:archi}, column 2). 
After a short grid search in a pilot experiment, we set $d$ to 128.
The embeddings are L2-normalized to the unit sphere.
The margin parameter $\alpha$ of the triplet loss is set to 0.5 and the learning rate to $10^{-6}$.
Each mini-batch contains 42 triplets of patches.
A given mini-batch represents one anchor track, one positive track and 42 negative tracks. 
Patches from these tracks are then randomly selected.

In the rest of the paper we denote this network as \textit{TL} .
We test it with the three sampling strategies presented in~\ref{ssec:mining}: \textit{Neighbors, Random uniform} and \textit{Distance-based} with 
$N_p$=15 and $N_n$=250.


\vspace{0.3cm}

{\bf Triplet loss system with Auto-pooling (TL Autopool)}:
In the \textit{AT Baseline} and \textit{TL} networks, the time dimension is progressively removed by a succession of max-pooling layers.
We test here the use of the Auto-Pooling layer proposed by McFee et al. \cite{McFee2018AdaptiveDetection}.
Two main adaptations were necessary to use Auto-pooling in our setup. 
First, the max-pooling sizes of the \textit{TL} network need to be  adapted to carry some temporal information until the last layer. 
Second, the number of filters needs to be divided by two due to GPU memory constraints. 
This network is refered to as \textit{TL Autopool} in the rest of the paper (see Table \ref{tab:archi}, column 3).
As for the \textit{TL} network, we train it to output embeddings using the triplet loss.
In the following, TL Autopool will only be tested with the Distance-based mining strategy.

\section{Evaluation of the proposed method}
\label{sec:eval}

\subsection{Dataset}
\label{sec:dataset}

To test our proposal, we need a dataset for which tags have been annotated at the track level and a similarity metric has been designed.
Such datasets exist in streaming services such as Pandora.
In our case, we use an extract of $N=14,246$ tracks from the catalog of Creaminal, a music supervision company.
This dataset is private and cannot be shared but the proposals made here are not specific to this dataset and can be applied to other ones.

Each track has a duration comprised between 45s and 5 minutes, and is sampled at 44100 Hz.
The taxonomy used for this dataset is made of $m=488$ tags, organized in 5 categories: Genre (e.g. Blues, Reggae, Electro-funk, Japanese Pop), Recording (e.g. Acoustic, Saturated, Guitar bass), Mood (e.g. Epic, Dancing, Nostalgic), Movement (e.g. Acceleration, Repetitive), and Lyrics (e.g. Death, Freedom, Nature). 
Each track was professionally annotated with a number of tags comprised between 5 and 35, the average number of tags per track being 16.8.
The dataset features a majority of Pop tracks, along with Electro, Rock, Country and Movie soundtracks.
The function $\mathcal{S}$ is also specific to this dataset and relies on a non-linear combination of weighted tags.
For our experiments, we split the dataset into a training, validation and test sets (60\%, 20\% and 20\% respectively)\footnote{ 
The same artist cannot be in both the training and test set. However it can appear both as a query and recommendations at test time; which may bias the results.}.
The distribution of tags is approximately the same in training and test. 

\vspace{-0.3cm}
\subsection{Evaluation metrics}

To evaluate our systems, we used each track of the test set as a query and ask the systems to rank all the other tracks of the test set by decreasing similarity with the query.

Let $R^{\mathcal{S}}_k(t) = [r_1(t), \ldots, r_{r}(k)]$ be the list $R^{\mathcal{S}}(t)$, truncated at rank $k$. 
Without loss of generality, we consider here that the \textit{relevant} tracks to recommend for a given test query $t$ are the five first tracks of the ranking:  $R^{\mathcal{S}}_{5}(t)$.
Thus, for each target track $t$, we ask our system to retrieve the 5 relevant ground truth recommendations $R_5^{\mathcal{S}}(t)$ among its $k$ estimated recommendation $R_k^{\mathcal{\hat{S}}}(t)$. 
Here we set $k$ to 20.
We then use four of the evaluation metrics proposed by~  \cite{Urbano2013EvaluationSimilarity}:

\begin{itemize} [leftmargin=4mm, parsep=0cm, itemsep=0cm, topsep=0cm]
    \item \textit{\ac{MAP}}: evaluates if the relevant tracks appear in high position in $R^{\mathcal{\hat{S}}}_{k}(t)$;
    \item \textit{Recall@k}: indicates which proportion of the relevant tracks appear in $R^{\mathcal{\hat{S}}}_{k}(t)$;
    \item \textit{Reciprocal rank (RR)}: is the inverse of the rank of the first relevant track in $R^{\mathcal{\hat{S}}}_{k}(t)$. Since we consider only the top $k$, the reciprocal rank is set to 0 if the rank is higher than $k$;
    \item \textit{\ac{nDCG}}: this metric allows to have a \textit{relevance scale} instead of binary relevance judgments (e.g., recommending $r_1(t)$ will produce a higher score than recommending $r_5(t)$ at the same rank in $R^{\mathcal{\hat{S}}}_{k}(t)$).
\end{itemize}



\vspace{-0.3cm}
\subsection{Results and discussion}

In Table~\ref{tab:res}, we compare the systems AT Baseline, TL (with the three mining strategies \emph{Neighbors}, \emph{Random uniform} and \emph{Distance-based}) and TL Autopool.
We indicate for each the average and confidence interval at 95\% of the four metrics (expressed as percentages).

\begin{table}[t!]
\small
\tabcolsep=0.142cm
  \begin{tabularx}{\columnwidth}{| l | c | c | c | c |}
    \hline
    \textbf{Model} & \textbf{MAP@20} & \textbf{Recall@20} & \textbf{RR@20} & \textbf{nDCG@20} \\ \hline \hline 
    \multirow{2}{*}{\textit{AT Baseline}}& 4.50 & 12.57 & 15.62 & 11.30 \\ 
    & $\pm$ 0.34 & $\pm$ 0.66 & $\pm$ 1.07 & $\pm$ 0.69 \\ \hline \hline 
    \multirow{2}{*}{\textit{TL Neighbors}} & 5.58 & 12.73 & 19.18 & 13.41 \\ 
    & $\pm$ 0.41 & $\pm$ 0.67 & $\pm$ 1.23 & $\pm$ 0.80 \\ \hline 
    \textit{TL Random} & 5.39 & 15.01 & 17.86 & 13.50 \\
    \textit{uniform} & $\pm$ 0.38 & $\pm$ 0.70 & $\pm$ 1.12 & $\pm$ 0.76 \\ \hline 
    \textit{TL Distance-} & \textbf{5.98} & \textbf{15.79} & \textbf{19.89} & \textbf{14.41} \\
    \textit{based}& \textbf{$\pm$ 0.40} & \textbf{$\pm$ 0.73} & \textbf{$\pm$ 1.19} & \textbf{$\pm$ 0.78} \\ \hline \hline 
    \textit{TL Autopool} & \textbf{7.99} & \textbf{17.74} & \textbf{24.68} & \textbf{17.95} \\ 
     \textit{(Distance-based)} & \textbf{$\pm$ 0.51} & \textbf{$\pm$ 0.79} & \textbf{$\pm$ 1.34} & \textbf{$\pm$ 0.92} \\ \hline  
  \end{tabularx}
  \caption{Comparison of the results of the AT Baseline vs TL systems (with various sampling strategies) vs TL Autopool. Higher is better.}
  \label{tab:res}
\end{table}

We first observe that the AT baseline is outperformed by all TL systems on all metrics.
This shows that in our case, learning the ranking directly is more efficient than learning the tags and applying $\mathcal{S}$ to their estimates. 
For information, the AT Baseline system (which replicates \cite{Choi2016AutomaticNetworks} architecture) achieves a mean-over-tag AUC of 0.79 on its auto-tagging task.

We then see that among the various mining strategies of the TL systems, the Distance-based negative sampling performs best on all metrics.
It should be noted that the Distance-based negative sampling was initially proposed by \cite{Manmatha2017SamplingLearning} which uses the learned embeddings to compute the distance online.
In our case, the distance can be calculated offline since we use the ground truth $\mathcal{S}$ instead of the embeddings for the distance computation. 
The Neighbors and Random uniform sampling strategies have similar performances, but the first has a better recall while the latter has a better reciprocal rank.

The last row of Table \ref{tab:res} indicates the results of the TL Autopool system (with Distance-based mining). 
We observe a boost in performances due to the added flexibility of Auto-pooling. 
This system is able to retrieve one of the 5 relevant tracks with almost a probability of 1/5 (Recall@20 = 17.74).
Note that in our case, neither the query nor the reference tracks have been seen during training.
The first relevant track is on average at rank 5.6 (inverse of RR@20=24.68), among approximately 2,900 test tracks. 
This makes our system a promising approach to efficient music retrieval in large datasets. 
Additionally, an informal listening test reveals that some of the recommended tracks, although judged "irrelevant" by our evaluation system, actually share important characteristics with the target.

\textbf{Reproducibility:}
Although we cannot distribute our private dataset and its oracle similarity function, to allow reproducibility of our work we provide the architecture and experimental code
\footnote{\href{https://gitlab.com/creaminal/publications/icassp2020-learning-to-rank-music-tracks}{https://gitlab.com/creaminal/publications/icassp2020-learning-to-rank-music-tracks}}.


\section{Conclusion and perspectives}
\label{sec:conclusion}

We propose here a method to learn a similarity ranking using a triplet loss network and a dataset of reference rankings.
We show that using the triplet loss to learn the ranking gives better results than learning the tags used by the ground truth similarity function.
This result is consistent across all our metrics. 
Finally, we show that Auto-pooling, proposed in the framework of audio event detection, also allows improvement in the case of music similarity.

Future works will focus on improving the training efficiency by mining on the fly useful triplets~\cite{Hermans2017InRe-Identification} \cite{Mishchuk2017WorkingLoss} \cite{Doras2019COVEREMBEDDINGS}.






\bibliographystyle{IEEEbib}
\bibliography{refs}

\end{document}